\documentclass[aps,prl,superscriptaddress,showpacs]{revtex4}
\usepackage{graphicx}
\usepackage{amsmath}
\usepackage{color}

\begin{document}

\title{Coherent meta-materials and the lasing spaser}

\author{N. I. Zheludev}
\homepage{www.nanophotonics.org.uk/niz}

\affiliation{Optoelectronics Research Centre, University of
Southampton, SO17 1BJ, UK}

\author{S. L. Prosvirnin}
\affiliation{Institute of Radio Astronomy, National Academy of
Sciences of Ukraine, Kharkov, 61002, Ukraine}

\author{N. Papasimakis}
\affiliation{Optoelectronics Research Centre, University of Southampton, SO17 1BJ, UK}

\author{V. A. Fedotov}
\affiliation{Optoelectronics Research Centre, University of
Southampton, SO17 1BJ, UK}

\date{\today}

\begin{abstract}

\end{abstract}

\maketitle

\textbf{In 2003 Bergman and Stockman introduced the spaser, a
quantum amplifier of surface plasmons by stimulated emission of
radiation \cite{stock03}. They argued that, by exploiting a
metal/dielectric composite medium, it should be possible to
construct a nano-device, where a strong coherent field is built up
in a spatial region much smaller than the wavelength
\cite{stock03, stock04}. V-shaped metallic inclusion, combined
with a collection of semiconductor quantum dots were discussed as
a possible realization of the spaser \cite{stock03}. Here we
introduce a further development of the spaser concept. We show
that by combining the metamaterial and spaser ideas one can create
a narrow-diversion coherent source of electromagnetic radiation
that is fuelled by plasmonic oscillations. We argue that
two-dimensional arrays of a certain class of plasmonic resonators
supporting high-Q \emph{coherent} current excitations provide an
intriguing opportunity to create spatially and temporally coherent
laser source, the Lasing Spaser.}

\
\
\begin{figure}[h]
\includegraphics[width=70mm]{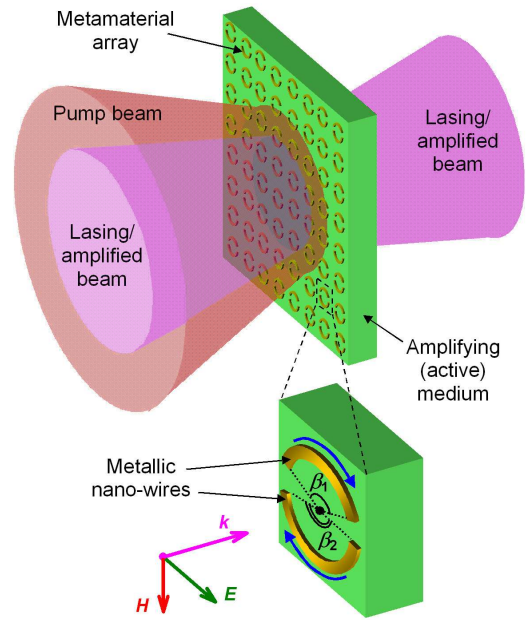}
\caption{The Lasing spaser consists of a gain medium slab (green)
supporting a regular array of metallic asymmetrically-split ring
resonators. The dashed box indicates an elementary translation
cell of the array, while the arrows along the arcs of the ring
illustrate the antisymmetric currents of plasmonic oscillations.
In-phase plasmonic oscillation in individual resonators leads to
the emission of spatially and temporarily coherent light
propagating in the direction normal to the array.}
\end{figure}

In the Lasing Spaser, identical plasmonic resonators impose the
frequency at which the device will lase. They will draw energy
from a supporting gain substrate. This ensemble of artificial
\emph{classical} electromagnetic resonators plays the role of the
active medium in the lasing spaser, just as an assemble of
essentially \emph{quantum} inversely populated atoms plays the
same role in a conventional laser. In the conventional laser the
direction of emission is imposed by the external resonator, while
its coherency is underpinned by the boson statistics of stimulated
emission of atoms in the gain medium. In the lasing spaser the
direction of emission is normal to the plane of the array. Here,
strong trapped-mode currents in plasmonic resonators will
oscillate in phase. However, the reason for coherence is not in
the boson statistics of stimulated emission, but the fact that an
in-phase collective oscillation of currents has the lowest losses
and is therefore the easiest to excite. A small asymmetry in the
plasmon resonator, which breaks the non-radiating nature of the
trapped mode oscillation, will allow a fraction of the energy
accumulated in current oscillations to be emitted by the spaser
array into the free space. This is analogous to the leakage of
radiation though the output coupler of a laser resonator.
Therefore in contrast with the optical quantum generator, the
Lasing Spaser is a classical device at all key levels apart from
the provision of gain to the substrate active medium.

To create a Lasing Spaser a special type of metamaterial array of
plamon resonators is needed. It should support high-Q current
oscillations that have lowest total emission losses when all
currents in the array oscillate in-phase. We will such media
\emph{coherent metamaterials}. We recently demonstrated that a
high-quality mode of intense antisymmetric current oscillations
may be excited in split ring resonators with weak asymmetry (ASRs)
\cite{ASR}. Strong oscillations in the rings will build up and
exhibit long decay time \emph{only} if the ring asymmetry is weak
and resonators are arranged into a regular two-dimensional array.
This is because the radiation losses associated with the electric
and magnetic dipole emission of the oscillating antisymmetric
currents are \emph{cancelled} if the resonators are placed in an
infinite regular array. Thus, the high-Q resonator is formed not
by a single ASR plasmonic resonator, but by the entire array. Weak
coupling of this current mode to free space occurs only due to the
asymmetry in the split ring and may be controlled by design
(smaller asymmetry gives lower coupling and higher Q-factor).
Behavior of the weakly asymmetric split ring arrays is in sharp
contrast with that of conventional metamaterials where the
response is determined by the dipolar resonance of the individual
elements of the structure, radiation losses are strong and depends
weakly on their mutual interactions. We argue that \emph{laser
action} fuelled by trapped-mode spaser current oscillations could
be achieved by exploiting the coherent nature and high-Q of the
oscillations in an array of ASRs and result in light emission with
high spatial coherence.

If the array of resonators is in contact with a gain medium, for
instance when it is supported by a thin slab of gain material (see
Fig. 1), then by introducing modest levels of gain in this high-Q
system, radiation losses and Joule losses in the metal can be
overcome. Various gain media such as optically and electrically
pumped semiconductor structures with direct gap and quantum
cascade amplification mechanisms or rare earth and semiconductor
quantum dot doped dielectrics may be suitable for this purpose. We
show below that on reaching the threshold value of gain, the
intensity of the resonant wave reflected and transmitted through
the structure increases dramatically. By combining a thin layer of
a gain medium with a high Q-factor ASR array, it is possible to
achieve \emph{orders of magnitude enhancement} of single-pass
amplification in comparison with the amplification of the bare
gain medium layer.

We illustrate this concept by providing numerical analysis of
amplification in the array of ASRs combined with a gain dielectric
substrate. Two cases are considered: In the first case, resonant
amplification is achieved in the mid-infrared part of the spectrum
(at a wavelength of about 8 $\mu m$), where Joule losses in the
metals can be neglected and only losses and gain in the isotropic
dielectric substrate are taken into account. In the second case,
we consider amplification at a wavelength of $1.65 \mu m$ and take
into account Joule losses in the metallic wires. In both cases,
losses and gain in the substrate are assumed to be frequency
independent. This simplifying assumption is valid when the
metamaterial resonance is narrower than the gain line of the
substrate and inhomogeneous spectral hole burning is
insignificant. We also assume no depletion of gain in all
operational regimes.

The unit cell of the modeled metamaterial structures is presented
in Fig.~1. It consists of a planar sub-wavelength asymmetric
metalic split-ring resonator (ASR) horizontally split in two wire
segments of different lengths corresponding to arc angles
$\beta_1$ and $\beta_2$, which are separated by equal gaps. The
ASR-resonator is brought into direct contact with a dielectric
slab, which could be a gain medium supporting the array. Arrays of
such metal structures can be manufactured by e-beam write and
photo lithography.

\begin{figure}
\includegraphics[width=80mm]{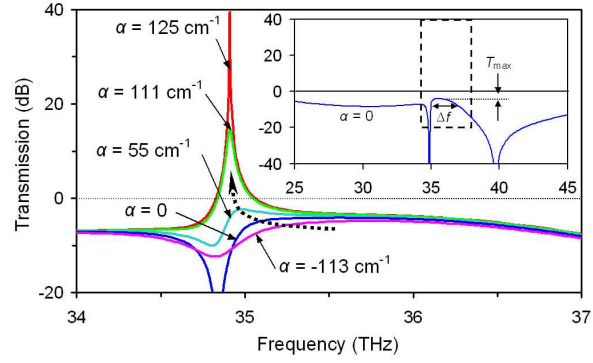}
\caption{Transmission spectra of the mid-IR planar
ASR-metamaterial in the vicinity of the trapped-mode transmission
resonance for different values of gain $\alpha$. The dashed arrow
follows the transformation of the transmission resonance. The
inset presents the transmission spectrum of the metamaterial with
no losses/optical gain in a much wider frequency range, while the
dashed box indicates the spectral domain that is covered by the
main plot.}
\end{figure}

Figure 2 shows the transmission characteristics of the infrared
ASR array for different levels of gain presented in term of the
gain coefficient $\alpha$. For negative values of $\alpha$ (lossy
substrate) the metamaterial attenuates electromagnetic radiation.
Gain in the substrate exceeding $\alpha_{th} = 70~cm^{-1}$ is
sufficient to overcome losses at a frequency of about $35~THz$
($\lambda=8.4~\mu m$) and signal attenuation becomes signal
amplification (see Fig.~3). This threshold level $\alpha_{th}$
corresponds to amplification of only $\sim 2.7\%$ in a $2~\mu m$
thick active layer of the bare substrate. A further increase in
the substrate gain leads to a rapid increase in resonant
amplification in the metamaterial reaching the level of $42~dB$
(approximately $\mu=1.6 \times 10^4$ times) at $\alpha =
125~cm^{-1}$. In a bare film such levels of gain will only lead to
amplification of about $5\%$. Alongside the increase in gain, the
width of the  amplified spectrum collapses from $1200~THz$ at zero
gain to $\Delta \nu =2~GHz$ at the amplification maximum. Further
increases in gain lead to a rapid decrease in amplification. This
is because gain broadens the amplification resonance in the same
way that losses broaden absorbtion resonances, and achieving
anti-phase oscillation of currents in the split ring arcs of the
plasmonic resonator becomes more difficult as radiation losses
increas.

\begin{figure}
\includegraphics[width=80mm]{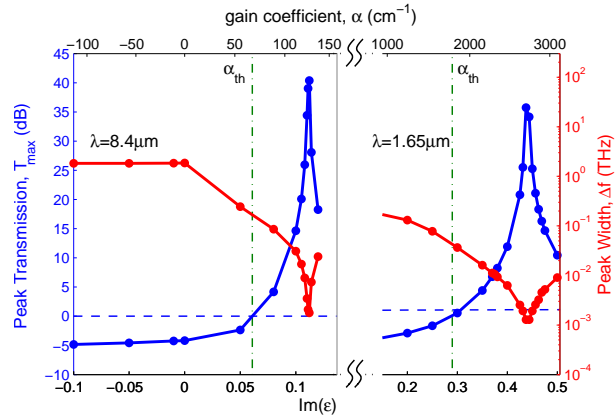}
\caption{Amplification (blue) and spectral width (red) of the
resonant transmission peak in both near- and mid-IR
ASR-metamaterials as functions of gain in the substrate. }
\end{figure}

\begin{figure}
\includegraphics[width=80mm]{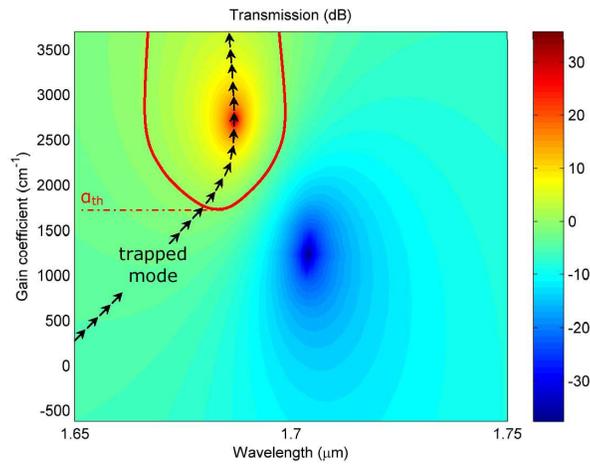}
\caption{Transmission spectra of the near-infrared
ASR-metamaterial in the vicinity of the trapped-mode transmission
resonance corresponding to different values of gain, $\alpha$.
Solid contour: region of unit transmission. The arrow line shows
evolution of the trapped mode resonance frequency with increase of
the gain.}
\end{figure}

Similar analysis has also been performed for a structure
resonating at $1.65\mu$, where losses in the metal increase the
threshold gain $\alpha_{th}$ to $\approx 1800~cm^{-1}$ and reduce
the maximum level of achievable wave amplification to about
$35~dB$ (approximately $\mu= 3.2 \times 10^3$ times) (see Figs.~3
and 4). Here amplification peaks at a gain level of about $\alpha
= 2550~cm^{-1}$, which corresponds to amplification of $5.5 \%$ in
the bare substrate film. Here the spectral width of the
amplification resonance reduces from $3~THz$ to about
$\Delta\nu=500~GHz$ at the optimal level of gain and maximum
amplification.

We argue that since the radiation losses in the metamaterial are
at a minimum when all currents in individual oscillators oscillate
coherently at the resonance frequency, the current oscillation
will self-start coherently in all rings of the array if sufficient
gain is provided. Such oscillations will produce a spatially and
temporally coherent diffraction-limited beam of optical emission
normal to the array, transforming optical amplifier into lasing
spaser. This will happen without the need for an external
resonator: coherence and narrow-diversion of the output will be
ensured by the low-loss condition. From the properties of the
metamaterial array as an amplifier we can expect that on reaching
a threshold gain the system will start lasing coherently across
the whole array. With increasing gain the output intensity will
increase rapidly while its spectrum will narrow dramatically. In
reality, the output intensity of the lasing spaser is likely to be
limited by saturation in the gain medium and heat management
problems.

The small scattering losses of the current in the metamaterial
array make the levels of threshold gain and gain needed to achieve
peak amplification of 35-40dB practically attainable. Indeed it
was recently demonstrated that quantum well structures can provide
high values of gain of the order of $10^3~cm^{-1}$ \cite{qwell},
which is similar to the threshold value required for an ASR array
operating at $1.65~\mu m$. Furthermore, quantum cascade amplifiers
can readily provide the gain values needed in the mid-infrared
case, since attainable gain coefficients in this wavelength range
exceed $100~cm^{-1}$ \cite{qc}. This easy-to-achieve threshold
gain condition gives a crucial advantage over the recent
suggestions to combine amplifying media with nano-shell
\cite{ziolk07} and horseshoe resonant elements \cite{sar} to
create a compact plasmonic nanolaser which is much smaller than
the wavelength. There the high dipole radiation losses of
plasmonic resonator make the threshold gain level extremely
difficult to achieve.

The lasing spaser allows high amplification and lasing in a very
thin layer of material with a modest gain level, making it a very
practical proposition. The thin-layer geometry is a desirable
feature for some highly-integrated devices and from the point of
view of heat management and integration. Here the
amplification/lasing frequency is determined by size of the ring
and may be tuned to match luminescence resonances in a large
variety of gain media such as rare-earth elements, quantum-cascade
amplifying media and quantum dots. All together this makes the
lasing spaser a generic concept for many applications.
\
\

\begin{acknowledgments}
The authors would like to acknowledge the financial support of the
EPSRC (UK).

\end{acknowledgments}

\
\

\textbf{ Method}

\
\

In the mid-IR version of the planar metamaterial the unit cell has
a lateral dimension of 1.5~$\mu m$, while the split ring has a
radius and line width of 0.6 and 0.05~$\mu m$ respectively, and
$\beta_1 = 160^{o}$, $\beta_2 = 151^{o}$. The thickness of the
active layer on the support substrate is 2~$\mu m$ and its
dielectric constant (real part) $\epsilon' = 10.9$. The optical
response of such metamaterial structure was analyzed in the
20-50~$THz$ frequency range (6 - 15~$\mu m$) using the method of
moments. This numerical method involves solving an integral
equation for the surface currents induced in the metallic pattern
by the incident electromagnetic wave, then calculating the
scattered fields produced by the currents as a superposition of
partial spatial waves. The metallic pattern is therefore treated
as a very thin perfect conductor (which is acceptable for most
metals in the mid-IR), while the gain (losses) in the substrate is
introduced through the imaginary part of its dielectric constant
and assumed to be isotropic.

In the metamaterial structure designed for the near-IR domain, the
diameter of the ASR-resonator is 140~$nm$, with a unit cell of
$210 \times 210$~$nm$. The angular lengths of the metallic wire
segments correspond to angles $\beta_1 = 160^{o}$, $\beta_2 =
125^{o}$ , while the size of their cross-section is $20 \times
50$~$nm$. The metal of the nano-wires is assumed to be silver with
a dielectric constant described by the Drude model. The substrate
is 100~$nm$ thick with $\epsilon' = 9.5$, and gain is introduced
through the imaginary part $\epsilon''$ of the substrate's
dielectric constant, which is related to the gain/attenuation
coefficient $\alpha$ by $\frac {2\pi}{\lambda}
Im(\sqrt{\epsilon'+i\epsilon''})$. The transmission properties of
this active nano-structure were numerically modeled in the
$500~nm~-~3~\mu m$ wavelength range using a true 3D finite element
method for solving Maxwell's equations, which also enabled us to
study the effect of gain anisotropy.


\end{document}